\documentclass[aps,prl,twocolumn,superscriptaddress]{revtex4-1}
\usepackage{graphicx}%
\usepackage{float}
\usepackage[T1]{fontenc}
\usepackage{color}
\usepackage[colorlinks]{hyperref}
\usepackage{amsmath}
\usepackage{amssymb}
\usepackage{bm}

\begin{document}
\title{Zeeman Splitting and Inverted Polarization of Biexciton Emission in Monolayer WS$_2$}


\author{Philipp Nagler}
\email{philipp.nagler@ur.de}
\affiliation{Department of Physics, University of Regensburg, D-93040 Regensburg, Germany}
\author{Mariana V. Ballottin}
\affiliation{High Field Magnet Laboratory (HFML - EMFL), Radboud University, 6525 ED Nijmegen, The Netherlands}
\author{Anatolie A. Mitioglu}
\affiliation{High Field Magnet Laboratory (HFML - EMFL), Radboud University, 6525 ED Nijmegen, The Netherlands}
\author{Mikhail V. Durnev}
\affiliation{Ioffe Institute, 194021 St. Petersburg, Russia}
\author{Takashi Taniguchi}
\affiliation{National Institute for Materials Science, Tsukuba, Ibaraki 305-004, Japan}
\author{Kenji Watanabe}
\affiliation{National Institute for Materials Science, Tsukuba, Ibaraki 305-004, Japan}
\author{Alexey Chernikov}
\affiliation{Department of Physics, University of Regensburg, D-93040 Regensburg, Germany}
\author{Christian Sch\"uller}
\affiliation{Department of Physics, University of Regensburg, D-93040 Regensburg, Germany}
\author{Mikhail M. Glazov}
\affiliation{Ioffe Institute, 194021 St. Petersburg, Russia}
\author{Peter C. M. Christianen}
\affiliation{High Field Magnet Laboratory (HFML - EMFL), Radboud University, 6525 ED Nijmegen, The Netherlands}
\author{Tobias Korn}
\email{tobias.korn@ur.de}
\affiliation{Department of Physics, University of Regensburg, D-93040 Regensburg, Germany}

\begin{abstract}

We investigate the magnetic-field-induced splitting of biexcitons in monolayer WS$_2$ using polarization-resolved photoluminescence spectroscopy in out-of-plane magnetic fields up to 30\,T. The observed $g$ factor of the biexciton amounts to $-3.89$, closely matching the $g$ factor of the neutral exciton. The biexciton emission shows an inverted circular field-induced polarization upon linearly polarized excitation, i.e. it exhibits preferential emission from the high-energy peak in a magnetic field. This phenomenon is explained by taking into account the configuration of the biexciton constituents in momentum space and their respective energetic behavior in magnetic fields. Our findings reveal the critical role of dark excitons in the composition of this many-body state.  

\end{abstract}

\maketitle
Monolayer transition metal dichalcogenides (TMDCs) are a fascinating platform to study the physics of Coulomb-correlated quasiparticles in the two-dimensional limit. Due to reduced dimensionality and dielectric screening, excitons in these materials possess binding energies on the order of 0.5\,eV, making them stable at room temperature and dominate the optical response \cite{He2014a,Ugeda2014,Chernikov2014,Ye2014a}. 
More recently, experimental evidence for biexcitons, where two excitons bind to a four-particle state has been brought forward in molybdenum- and tungsten-based monolayer TMDCs \cite{You2015,Shang2015,Plechinger2015b,Sie2015,Sie2016,Lee2016b,Kim2016,Okada2017,Paradisanos2017,Hao2017,Pei2017}. These excitonic molecules are subject to intriguing many-body physics and could serve as a platform for future quantum optics experiments due to their cascaded emission accompanied by entangled photon generation \cite{He2016c}. 
However, key questions with respect to the nature of the biexciton \cite{Tuan2017} and specifically the composition of this many-body state in momentum space remain open. In this respect, probing the behavior of excitonic complexes in strong magnetic fields has proven to be a powerful tool to gain a detailed understanding of the properties of these quasiparticles. Recently, this approach has revealed fundamental insights into neutral and charged excitons of atomically thin TMDCs \cite{Aivazian2014,Li2014,Srivastava2015,Macneill2015a,Wang2015,Mitioglu2015a,Stier2016,Plechinger2016}. Moreover, it was shown that the observed magnetic-field-induced population imbalances can be crucial for drawing conclusions on the composition of many-body states \cite{Astakhov2005,Bartsch2011}. Therefore, we expect to learn critical information such as the $g$ factor and the field-induced polarization by probing biexcitons in an atomically thin semiconductor in high magnetic fields.

Here, we investigate the properties of biexcitons in monolayer WS$_2$ under the influence of an external out-of-plane magnetic field up to 30\,T. The magnetic field lifts the valley degeneracy of the biexciton, allowing us to extract its spectroscopic $g$ factor of $-3.89$, in close agreement to the spectroscopic $g$ factor of the neutral exciton of $-3.82$ and thus providing further evidence for the concept of biexcitons in atomically thin TMDCs. Under linearly polarized excitation, we observe an inverted polarization of the biexciton emission in the magnetic field, implying that the state that emits at the higher energy is preferentially occupied, in contrast to the behavior of the neutral exciton. These observations, together with the theoretical analysis, allow us to draw conclusions on the valley configuration of the biexciton in WS$_2$, inferring that it consists of a bright exciton in one valley and an intra-valley dark exciton in the other valley. 

\begin{figure}
	\centering
	\includegraphics*[width=1\linewidth]{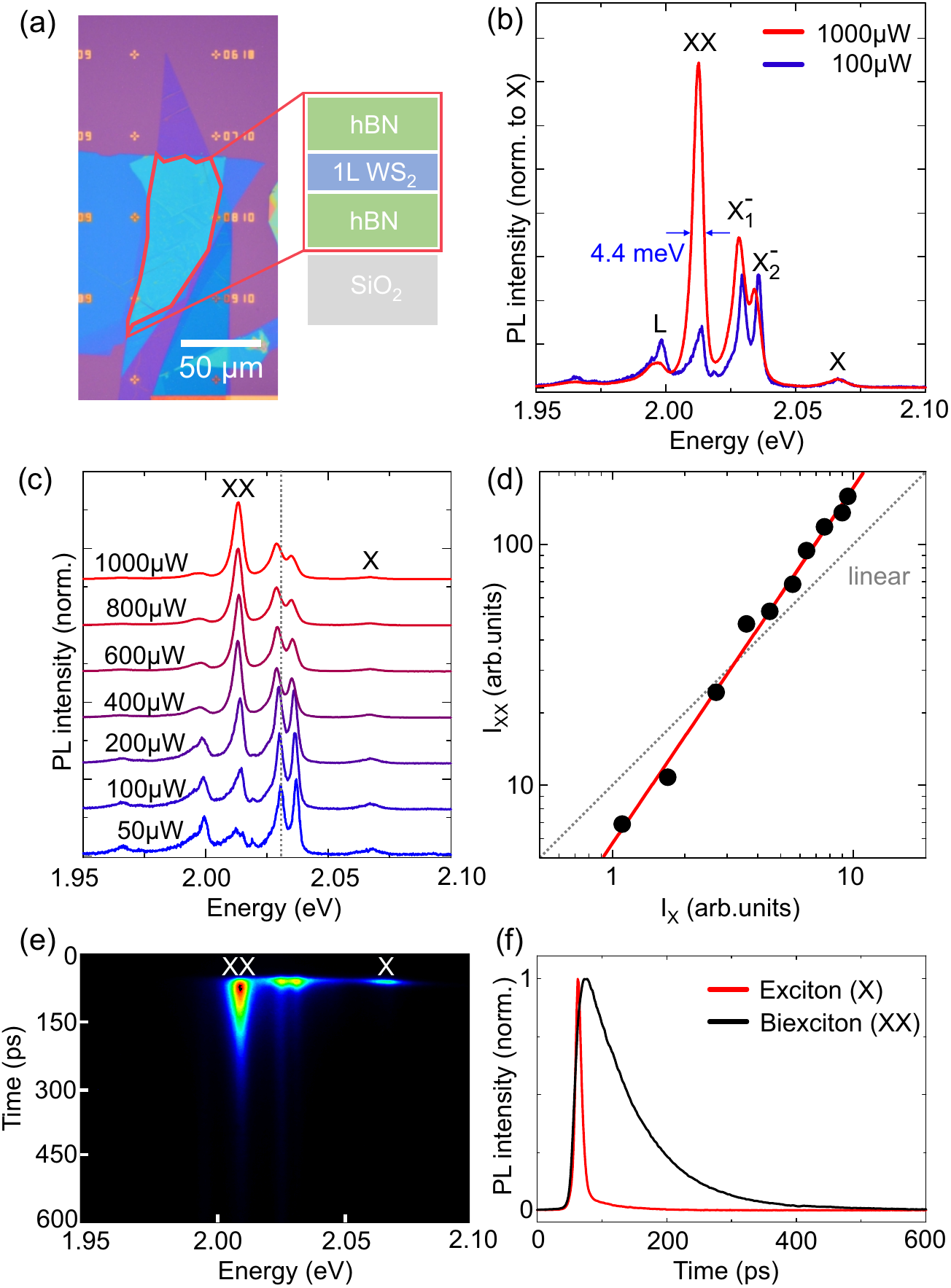}
	\caption{(a) Optical micrograph of the hBN/WS$_2$/hBN/SiO$_2$ heterostructure under study. (b) PL spectra at 4\,K taken on the heterostructure under excitation powers of 100\,$\mu$W (blue) and 1000\,$\mu$W (red). The spectra are normalized to the intensity of the neutral exciton (X). (c) PL spectra at 4\,K for the indicated applied excitation powers. The grey dashed line marks the energetic position of the low-energy trion state (X$_1^{-}$) at 50\,$\mu$W. The spectra are normalized to the peak with the highest intensity. (d) Double logarithmic representation of the intensity of the biexciton (XX) as a function of the intensity of the neutral exciton. The red line is a power-law fit with $I_{XX} \sim (I_{X})^{\alpha}$ with $\alpha=1.48$. The gray dashed line indicates a linear relation. (e) False-color plot of TRPL spectrum measured at 4\,K and 40~$\mu$W. (f) Normalized TRPL traces of the X and XX features extracted from (e).
	}
	\label{Plot1}
\end{figure}

The sample in this study (see Figure \ref{Plot1}(a)) was fabricated by an all-dry transfer technique \cite{Castellanos-Gomez2014a} and consists of a monolayer of WS$_2$ (bulk crystals from HQ Graphene) which is sandwiched between two thin sheets of hexagonal boron nitride (hBN). Static PL measurements were performed using excitation by a continuous-wave laser with a photon energy of 2.21\,eV. The laser was focused by an objective to a spot size of about 4\,$\mu$m. The reflected PL signal was collected by the same objective and measured by a spectrometer equipped with a liquid-nitrogen cooled CCD. All experiments presented in the main manuscript have been conducted at a nominal sample temperature of 4.5\,K. 

A characteristic photoluminescence spectrum of the structure at an excitation power of 100\,$\mu$W is shown in Fig. \ref{Plot1}(b) (blue line). In agreement with recent reports, the encapsulation of the monolayer TMDC with hBN results in significantly reduced linewidths of the excitonic features and thus drastically enhances the optical quality of the studied system \cite{Wang2016b,Wang2017,Cadiz2017,Manca2017,Ajayi2017,Wierzbowski2017}. In the spectrum we can clearly resolve the neutral exciton (X) at 2.067\,eV and the two trion species X$_1^{-}$ and X$_2^{-}$ at an energy of 2.030\,eV and 2.036\,eV which are split due to Coloumb exchange interaction \cite{Yu2014a,Jones2015,Plechinger2016a,Courtade2017a}. Furthermore, at an energy of 1.998\,eV we observe emission which is typically attributed to the recombination of carriers localized at defects (L) or alternatively was recently interpreted as phonon-assisted emission \cite{Lindlau2017a}.
The red line in Fig. \ref{Plot1}(b) depicts the spectrum with an excitation power of 1000\,$\mu$W and normalized to the intensity of the neutral exciton X. The intensity of the two trion features scales like the exciton intensity and the emission from the defect state L saturates. Most importantly, at this high excitation power, a novel feature emerges at 2.012\,eV exhibiting a clear super-linear behavior with respect to the other peaks. We attribute this peak to the emission of biexcitons (XX), in line with recent literature \cite{You2015,Shang2015,Plechinger2015b,Okada2017,Paradisanos2017}. The narrow linewidth of about 4.4\,meV of the biexciton further confirms the high sample quality.

The super-linear increase of the biexciton peak is illustrated in Fig. \ref{Plot1}(c) where spectra normalized to the intensity of the maximum peak for different excitation powers are depicted. Relating the intensity of the biexciton with the intensity of the neutral exciton by a power-law fit with $I_{XX} \sim (I_{X})^{\alpha}$ yields a factor $\alpha$ of 1.48, see Fig. \ref{Plot1}(d). While $\alpha=2$ would be expected for full thermal equilibrium between neutral exciton and biexciton, values for $\alpha$ smaller than 2 have been regularly observed recently in TMDCs and were linked to a lack of full equilibrium between the two states \cite{You2015}. 

To track the time-resolved dynamics of neutral exciton and biexciton we employ a streak camera combined with a frequency-doubled pulsed fiber laser system at an excitation energy of 2.21\,eV. The respective time-resolved traces are shown in Fig. \ref{Plot1}(e) and (f). While the decay of the neutral excitons is faster than our system resolution (10\,ps), the biexciton decay can be readily quantified. A monoexponential fit yields a decay time of 83\,ps for the biexciton. A possible contribution to a slower decay rate of the biexciton PL may be explained within the excitonic molecule model of Ref.~\cite{Citrin1994}, which predicts the reduction of the biexciton radiative decay rate as compared to the exciton rate roughly by a factor $\mu \sim (qa)^2 \ll 1$, where $q$ is the wave vector of light inside a monolayer and $a$ is the interexcitonic separation within the biexciton, see Supplementary Material for details.
The emission of the defect state L occurs on a far longer timescale than the biexciton, clearly differentiating the origin of the two species (see Supplemental Material for data). 

\begin{figure}
 	\centering
 	\includegraphics*[width=1\linewidth]{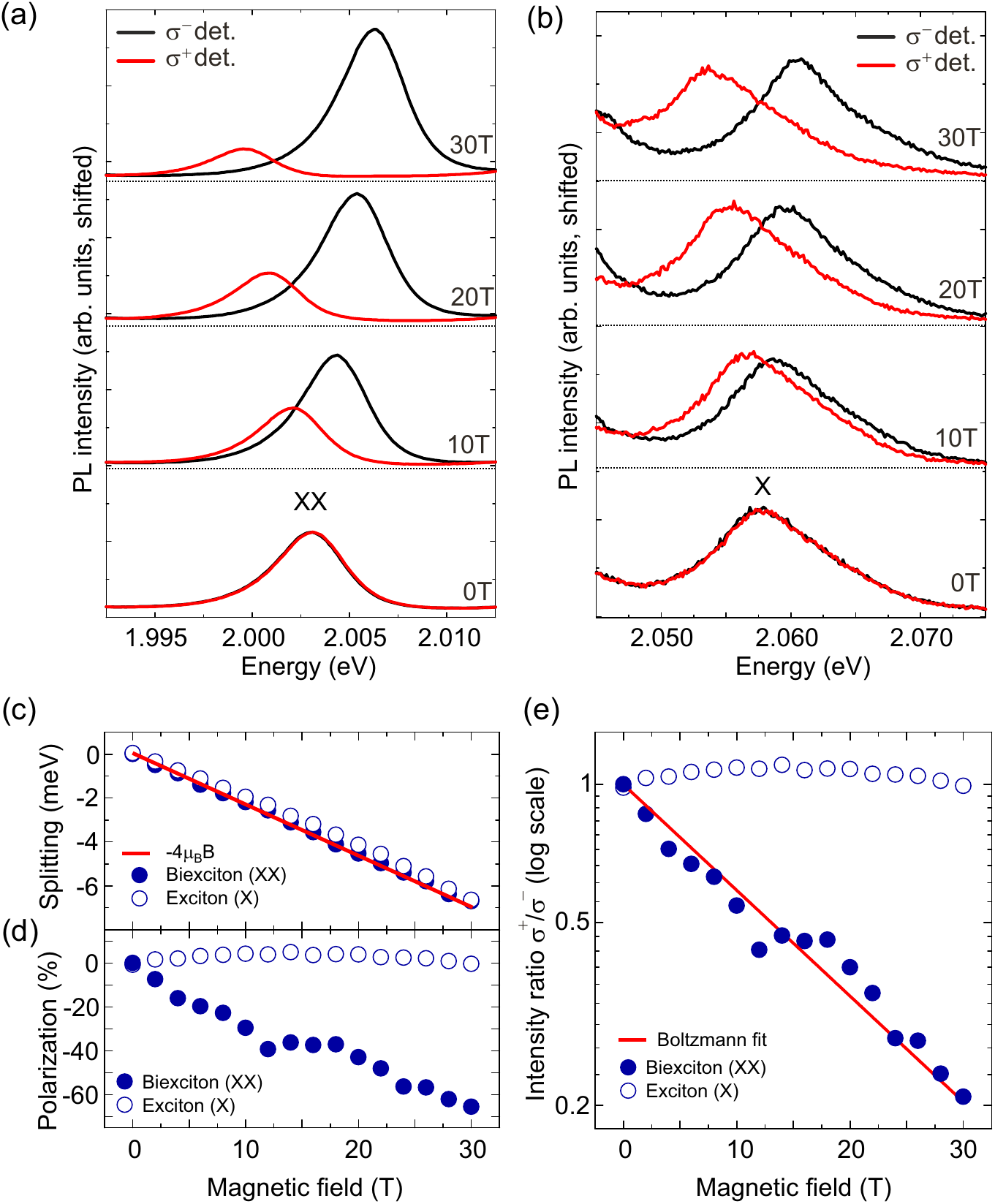}
 	\caption{(a) PL spectra of the XX peak for $\sigma^+$ and $\sigma^-$ polarized detection after excitation with linearly polarized light for various out-of-plane magnetic fields up to 30\,T. (b) Same as in (a) but for the X peak. (c) Peak splitting of the X peak (open circles) and the XX peak (filled circles) in dependence of magnetic field. The solid red line indicates a linear dependence with a $g$ factor of $-4$. (d) Magnetic-field-induced polarization of the exciton and the biexciton. (e) Corresponding intensity ratio (log. scale) of the $\sigma^+$ and $\sigma^-$ polarized light components in detection. The solid red line is a fit to a thermal Boltzmann distribution, Eq.~(\ref{Boltz}), with $g_{T}^{XX}=4$ and $T=50$~K. 
 	}
 	\label{Plot2}
 \end{figure}

We now turn to measurements of the biexciton and exciton spectra in an out-of-plane (Faraday configuration) magnetic field of up to 30\,T. All measurements in magnetic field have been carried out with a linearly polarized laser, populating both valley configurations equally. The emission is analyzed in a circularly polarized basis, which allows us to resolve the resulting magnetic splitting and to quantify the degree of polarization. Figure \ref{Plot2}(a) and (b) show a series of spectra of the XX and X peak at 0\,T, 10\,T, 20\,T and 30\,T for both detection polarizations. The field-induced energetic splitting for both exciton and biexciton is very similar and amounts to $\approx6.7$\,meV at 30\,T for both features. The resulting energetic splitting of the $\sigma^\pm$-polarized emission peaks with energies $E^{\sigma\pm}$ for exciton and biexciton is shown in Fig. \ref{Plot2}(c). Using the definition 
\begin{equation}
	\label{Lande}
	\Delta E = E^{\sigma^+}-E^{\sigma^-}=g_S\mu_{B}B,
\end{equation}
where $\mu_B\approx 58$~$\mu$eV/T is the Bohr magneton, $B$ is the magnetic field, and $g_S$ is the spectroscopic $g$ factor of the emitting state, we obtain $g_S^{\rm XX}=-3.89$ for the biexciton.
The deduced spectroscopic $g$ factor of the neutral exciton amounts to $g_S^{\rm X}=-3.82$, in very close agreement to previous measurements on WS$_2$ \cite{Stier2016,Plechinger2016,Kuhnert2016} and almost identical to that of the biexciton. The red line in Fig. \ref{Plot2}(c) marks a linear dependence with a $g$ factor of $-4$ and serves as a guide to the eye. 
 
However, the intensities of the circular components of emission behave distinctively different for the X and XX emission at high magnetic fields. 
The corresponding field-induced circular polarization, defined as $P=(I_{\sigma^+}-I_{\sigma^-})/(I_{\sigma^+}+I_{\sigma^-})$, is shown in Fig. \ref{Plot2}(d). For the exciton, neither polarization component is favored, i.e., its emission is mostly unpolarized, but for the biexciton the intensity of the energetically higher polarization component ($\sigma^-$ for $B>0$ and $g^{\rm XX}<0$) is strongly increased as compared with the lower-energy component and reaches values close to  $-70\%$ at 30\,T. A qualitatively similar depiction of this situation is given in Fig. \ref{Plot2}(e) where the ratio of the two intensities $I_{\sigma^+}/I_{\sigma^-}$ is plotted vs. the magnetic field, which will be discussed in more detail below. 

While the absence of circular polarization at the exciton resonance can be easily understood as a result of its very short, $\lesssim 10$~ps, lifetime, during which excitons may still be far from equilibrium, the inverted polarization of biexcitons is less obvious. We understand this result by recalling that the population behavior of an excitonic species in a magnetic field is determined by the change of the total energy of the state and not of the species that is responsible for the observed emission feature \cite{Astakhov2005,Bartsch2011}. To take this into account, we introduce a total $g$ factor $g_{T}$, which describes the Zeeman shifts of all constituents of the biexciton, i.e. the total energy of the composite exciton state. Hence, assuming quasi-equilibrium, the ratio of the two intensities $I_{\sigma^+}/I_{\sigma^-}$ that reflects the relative populations can also be expressed by means of a Boltzmann distribution as $I_{\sigma^+}/I_{\sigma^-}=\exp(-g_{T}\mu_{B}B/k_BT)$, where $T$ is the effective temperature of the biexciton gas and $k_B$ is the Boltzmann constant. An inverted polarization behavior can thus be expected if $g_{T}>0$. On the other hand, the spectroscopic $g$ factor $g_{S}$ only refers to the energies of the emitting state. Hence, for an exciton we always have $g_{S}=g_{T}$, while for more complex many-body states such as the biexciton, the total $g$ factor $g_{T}$ can differ from $g_{S}$ since it contains also the energies of the non-emitting states. Our analysis below shows that biexcitons are close to thermal equilibrium, however, the total energies of biexciton states are not directly related to the emission energies due to the presence of the second, dark exciton in this four-particle complex. 

\begin{figure}
	\centering
	\includegraphics[width=\linewidth]{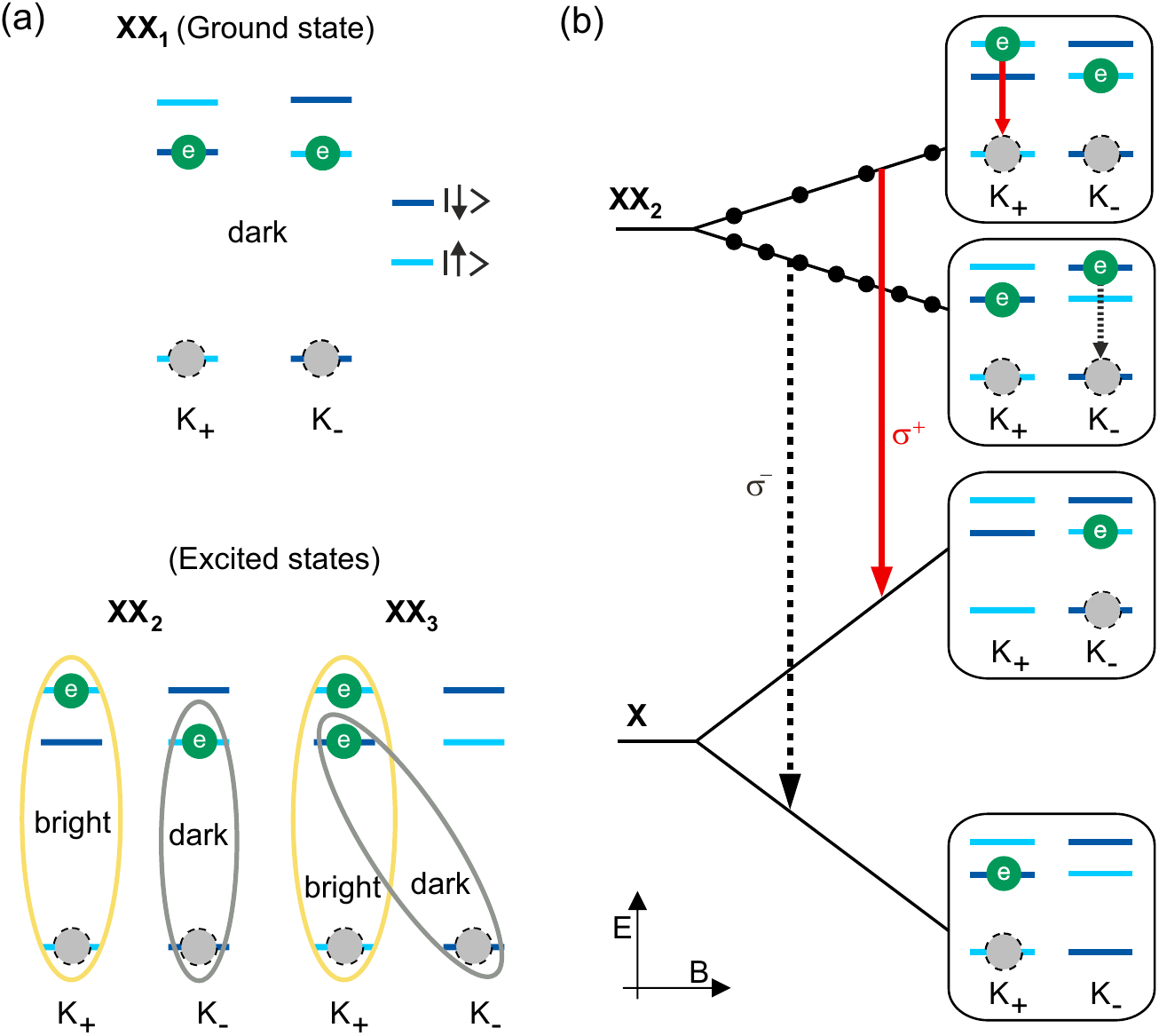}
	\caption{(a) Three configurations of biexcitons in monolayer WS$_2$. Both excited states, XX$_2$ and XX$_3$, have one transition which is both spin- and momentum-allowed. The remaining dark state is either spin-forbidden (XX$_2$) or indirect in momentum space (XX$_3$). Green circles denote electrons in the conduction band and grey circles denote unoccupied states in the valence band.
	(b) Schematics of biexciton XX$_2$ recombination which demonstrates the difference in the emission energies and the total biexciton energies. The number of dots sketches relative populations of the Zeeman-split XX$_2$ levels. Solid and dashed arrows denote recombination paths in $\sigma^+$ and $\sigma^-$ polarization, respectively.
	}
	\label{Plot3}
\end{figure}

To elucidate the expected polarization behavior of biexcitons in a magnetic field, let us first consider different configurations of the biexciton in WS$_2$ without magnetic field. Here, we adopt the conventional basic picture of the biexciton state in WS$_2$ as being charge-neutral and originating from two excitons in the $\bm K$ valleys. In the ground state XX$_1$, as shown in the upper part of Fig. \ref{Plot3}(a), the two electrons occupy the lowest-lying conduction bands of the $\bm K_+$  and $\bm K_-$ valleys, respectively, and the holes the highest-lying valence bands. While this arrangement is energetically the most favorable, it is optically dark since intra-valley electron-hole transitions are spin-forbidden in this case~\footnote{The electron-electron exchange interaction can mix this biexciton state with the excited one, XX$_4$, where two electrons occupy the highest conduction subbands~\cite{Courtade2017a,Danovich2017} and make this state optically active. One cannot expect sizable magneto-induced polarization of this XX$_1$ biexciton, see Supplemental Material for details.}. Thus, we do not detect this configuration in our optical experiments. The other two configurations in Fig. \ref{Plot3}(a), XX$_2$ and XX$_3$, consist of one bright electron-hole pair, which can recombine with emission of a photon, and one dark electron-hole pair, which is optically inactive. 

The Zeeman splitting of biexciton states can be analyzed by considering the corresponding shifts of the conduction and valence bands. The Zeeman Hamiltonian for an electron in the conduction, $c$, or valence band, $v$, reads
	\begin{equation}
	\mathcal H_{c,v} = \frac12 \mu_B B (g^{\rm orb}_{c,v} \tau_z + g^{\rm sp}_{c,v} \sigma_z)\:,
	\end{equation}
where $g^{\rm orb}$ is the $g$ factor describing orbital contributions to the Zeeman splitting ($\tau_z = \pm 1$ is the valley index), and $g^{\rm sp}$ describes the spin contribution ($\sigma_z = \pm 1$ is the spin index). Since the radiative recombination pathway of the biexciton states XX$_2$ and XX$_3$ is essentially identical to that of the optically bright exciton, the spectroscopic $g$ factor of these complexes, introduced in Eq.~(\ref{Lande}), is the same as of the bright exciton, and equals to
	\begin{equation}
	g_S^{\rm XX} = g_S^{\rm X} = g_c^{\rm orb} + g_c^{\rm sp} - g_v^{\rm orb} - g_v^{\rm sp}\:.
	\end{equation}
Using recently calculated values for WS$_2$~\cite{Rybkovskiy2017} with $g_v^{\rm orb} = 3.96$ and $g_c^{\rm orb}=0.11$ and $g_c^{\rm sp} = g_v^{\rm sp} = 2$ we obtain a theoretical value of $g^{\rm XX}_{th} = - 3.85$, in a good agreement with our experimental observation of $g_S^{\rm X}=-3.82$ and $g_S^{\rm XX}=-3.89$. Note that these estimates are also consistent with the approach of Refs.~\cite{Li2014,Aivazian2014,Srivastava2015}. 

The distribution of biexcitons over the Zeeman-split states is determined, however, by the total energy of the biexciton which accounts for, both, the energy of bright and dark exciton. Correspondingly, the total Zeeman splitting of a biexciton state, proportional to the total $g$ factor $g_T^{\rm XX}$ differs from the spectroscopic $g$ factor, $g_S^{\rm XX}$. Both XX$_2$ and XX$_3$ states consist of a pair of holes in a singlet state which is unaffected by the magnetic field and a pair of electrons with same spins located in different valleys (XX$_2$) or in the same valley but with opposite spins (XX$_3$) yielding

	\begin{equation}
	\label{g*}
	g_T^{\rm XX_2} = 2 g_c^{\rm sp}\approx +4\:,\:\:\: g_T^{\rm XX_3} = 2 g_c^{\rm orb}\approx +0.2\:,
	\end{equation}
where the estimates have been obtained taking $g_c^{\rm orb}$ from Ref.~\cite{Rybkovskiy2017}. In this case, where $g_T^{\rm XX}$ is positive, the $\sigma^-$-active biexciton state is energetically lower than the $\sigma^+$-active one, and hence is more populated leading to an inverted polarization. The expected evolution of energy levels and the corresponding population for this situation are depicted in Fig.~\ref{Plot3}(b). Naturally, one can also arrive at Eq.~(\ref{g*}) by summing up the Zeeman splittings of bright and dark excitons, see Supplemental Material for details. 

According to the estimates in Eq.~(\ref{g*}), both, XX$_2$ and XX$_3$ biexcitons lead to inverted polarization. In a simplified picture we can assume that biexcitons are distributed between the two states in a quasi-equilibrium during emission. Further insight into the origin of the emitting biexciton state is given by the fit of the experimental data in Fig.~\ref{Plot2}(e) with the following equation:

	\begin{equation}
	\label{Boltz}
	\frac{I_{\sigma^+}}{I_{\sigma^-}}=\exp{\left(-\frac{g_{T}^{\rm XX}\mu_B B}{k_B T}\right)}.
	\end{equation}
From this fit we can extract only the ratio of the total $g$ factor $g_T$ and the effective temperature of the biexciton gas $T$. Using theoretical estimations for $g_T$ given by Eq.~\eqref{g*} we obtain an effective temperature $T \approx 50$~K for the XX$_2$ complex and $T \approx 2.5$~K for the XX$_3$ complex. The fact that in the latter case the value of $T$ is lower than the nominal sample temperature in our setup of 4.5\,K strongly indicates that the inverted polarization is provided by the XX$_2$ biexciton.

The deviation of $T$ from the nominal sample temperature can stem either from the biexciton gas being out of equilibrium with the lattice due to slow cooling by acoustic phonons or from the heating of the crystal itself due to laser excitation with high power (1000\,$\mu$W). The latter effect can be estimated by determining the power-induced energetic redshift of the trion which amounts to 2.1\,meV at 1000$\,\mu$W, see dashed line in Fig. \ref{Plot1}(c). Comparing this shift to the typical temperature dependence of the peak energies of the sample with increasing temperature (see Supplemental Material for data) we estimate a lattice temperature of 35\,K, in qualitative agreement with the resulting temperature from the Boltzmann fit. 
 
In summary, we have measured the polarization-resolved photoluminescence of biexcitons and excitons in monolayer WS$_2$ in a perpendicular magnetic field up to 30\,T. The lifting of the valley degeneracy allows us to determine the spectroscopic $g$ factor of the biexciton $g_S^{XX}$ to be $g_S^{XX}=-3.89$, closely matching the spectroscopic $g$ factor of the exciton $g_S^{X}=-3.82$. The agreement of spectroscopic $g$ factors between the two different excitonic species gives additional evidence for the formation of biexcitons in monolayer TMDCs. Interestingly, we have observed that the sign of magneto-induced circular polarization does not match the sign of the Zeeman splitting of biexciton emission. This observation is explained by taking into account the evolution of the total energy of the biexciton in a magnetic field. On the basis of the experiment and developed model we are able to identify the optically dominant excited biexciton state of monolayer WS$_2$. Our results form a basis for future experiments on these four-particle states and highlight the importance of the dark states involved in the formation of biexcitons. 

Financial support by the DFG via GRK 1570, KO 3612/1-1, SFB 689, SFB 1277 (B05) and CH 1672/1-1 and support of HFML-RU/FOM, member of the European Magnetic Field Laboratory (EMFL) is gratefully acknowledged. MVD acknowledges financial support from RFBR project No. 16-32-60175 and the RF President grant MK-7389.2016.2. MMG was partially supported by RF President grant MD-1555.2017.2. Growth of hexagonal boron nitride crystals was supported by the Elemental Strategy Initiative conducted by the MEXT, Japan and JSPSKAKENHI Grant Numbers JP15K21722.
\newpage
\onecolumngrid
\appendix
\section{Supplementary Information: \\Zeeman Splitting and Inverted Polarization of Biexciton Emission in Monolayer WS$_2$}

\subsection{1.~Decay dynamics of the L peak in comparison to the XX peak}

\begin{figure}[hhhhh]
	\centering
	\includegraphics*[width=0.9\linewidth]{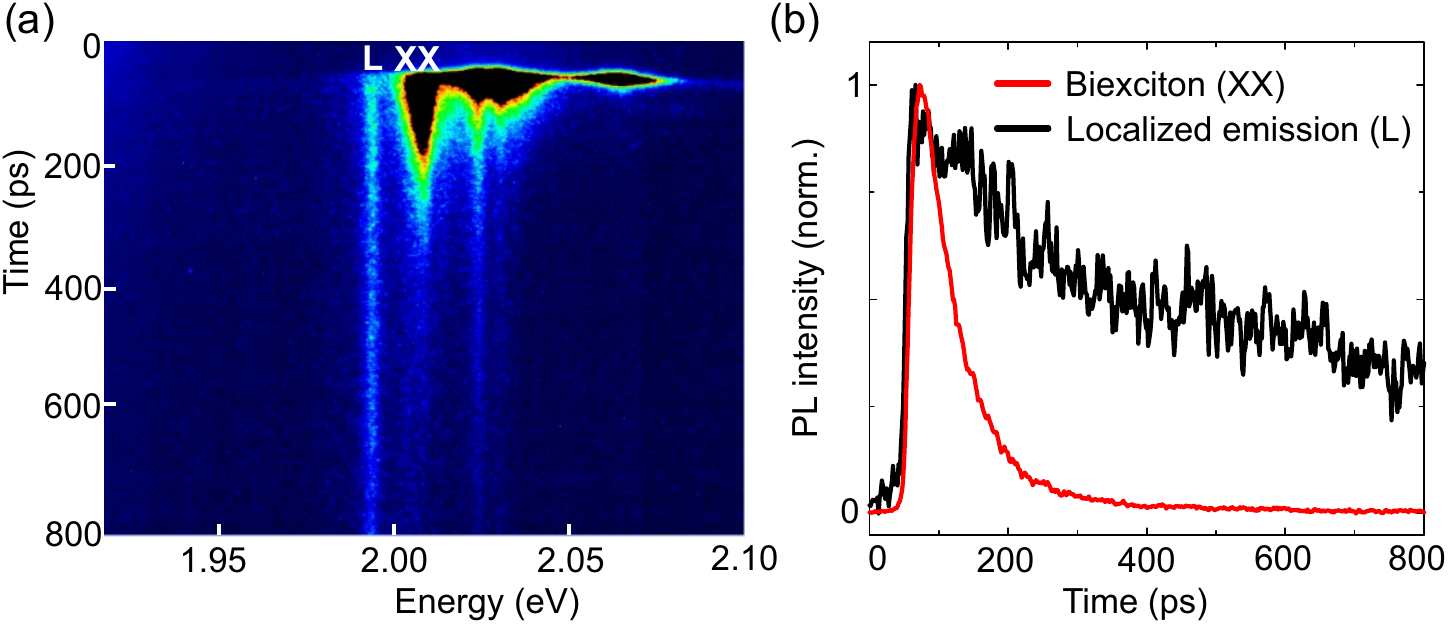}
	\caption{(a) Streak camera image of the decay dynamics at a nominal sample temperature of 4.5\,K. (b) Extracted normalized traces comparing the emission of localized defects (L) and biexcitons (XX). 
	}
	\label{S1}
\end{figure}

Figure \ref{S1}a shows the decay dynamics of the excitonic features at low excitation power (5\,$\mu$W) in a streak camera image. As can be already inferred from this image, the decay of the localized defects occurs on far longer timescales than the biexciton XX. 

These different decay dynamics are further illustrated in Fig. \ref{S1}(b) where individual traces of the two features are shown. Using a single exponential decay function we obtain a decay time of the biexciton of $\tau_0 = 58$\,ps and for the L peak a decay time of 256\,ps. Note that the value of $\tau_0$ measured here at an excitation power 5\,$\mu$W is smaller than the one mentioned in the main text, which was measured at 40\,$\mu$W excitation.

\clearpage

\subsection{2.~Power-induced redshift and actual sample temperature at 1000\,$\mu$W excitation power}

\begin{figure}[hhhhh]
	\centering
	\includegraphics*[width=0.7\linewidth]{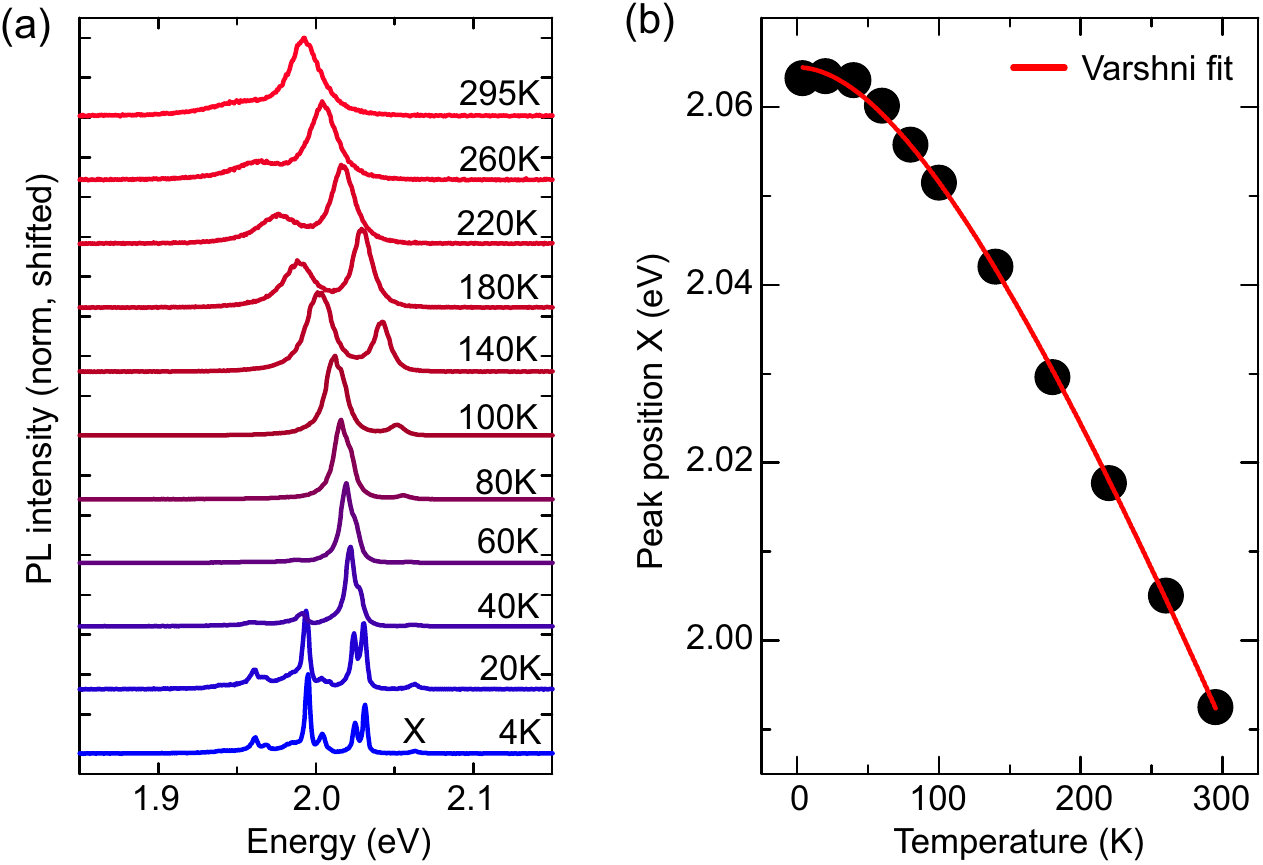}
	\caption{(a)Temperature-dependent PL spectra of the hBN/WS$_2$/hBN/ heterostructure at a constant excitation power of 5\,$\mu$W. (b) Peak position of the neutral exciton (X) in dependence of temperature. The red line is a Varshni fit to the data. 
	}
	\label{S2}
\end{figure}

In the following we discuss the power-induced heating of the sample due to the high excitation powers (1000\,$\mu$W). As already mentioned in the main text, we can estimate the sample temperature by comparing the power-induced redshift of an excitonic resonance (see Fig. 1(c) of main text) to the temperature-dependent shift which is described by a Varshni fit. 
Figure  \ref{S2}(a) shows temperature-dependent PL spectra obtained for the sample under study. The resulting evolution of the peak position of the neutral exciton X is shown in Fig.~\ref{S2}(b).

A phenomenological description of the shift of the band gap with respect to temperature is given by the Varshni equation:
\begin{equation}
\label{Varshni}
E_g(T)=E_g(0)-\frac{\alpha T^2}{T+\beta}  \hspace{4pt}.
\end{equation}
Fitting the data to Eq.~\eqref{Varshni} yields $E_g(0)=2.0644$\,eV, $\alpha=4.45\cdot10^{-4}$\,eV/K and $\beta=247.5\,$\,K. Using the induced redshift of 2.1\,meV, see Fig. 1(c) of the main text, and comparing it to the obtained Varshi fit formula we obtain an estimated sample temperature of 35\,K. 

\subsection{3.~Total $g$ factor of excited biexcitons in the excitonic picture}

\begin{figure}[h]
	\centering
	\includegraphics*[width=0.7\linewidth]{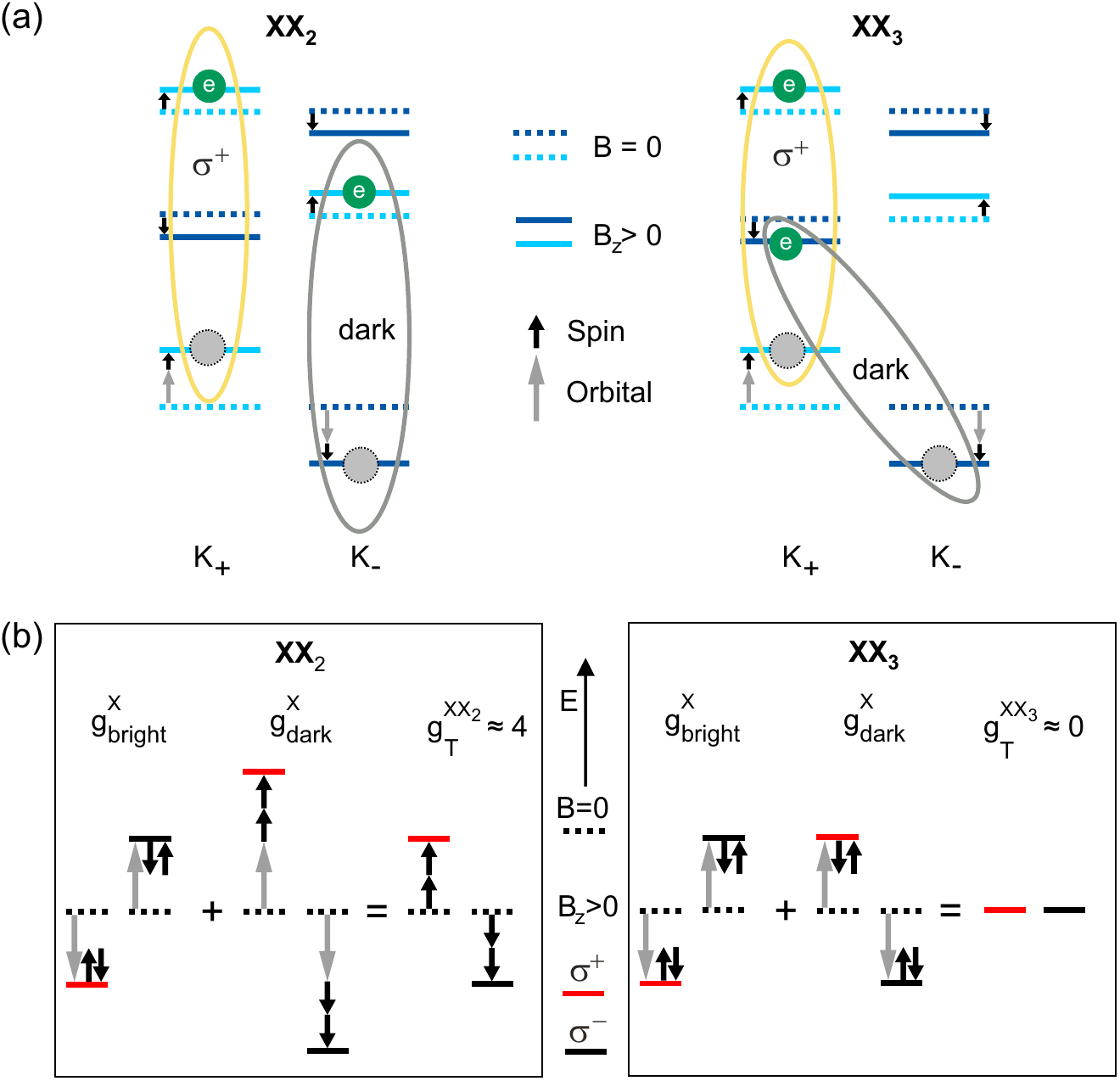}
	\caption{(a) Evolution of the energy levels for the $\sigma^+$ state of XX$_2$ and XX$_3$ with positive ($B_z>0$) applied magnetic field. Dashed lines indicate the situation for $B=0$.  Grey arrows depict the contribution from spin, black arrows the orbital contribution to the overall valley-selective splitting. (b) Energy diagram of the two excited states XX$_2$ and XX$_3$ with the respective $g$ factors of the bright exciton ($g^{\rm X}_{\rm bright}$), the dark state ($g^{\rm X}_{\rm dark}$)  and the resulting total $g$ factor of the biexciton ($g_T^{\rm XX}$). Dashed lines indicate the situation for $B=0$. Red (black) lines show the energy levels for $\sigma^-$ ($\sigma^+$) -polarized components in a positive magnetic field.
	}
	\label{S3}
\end{figure}

As mentioned in the main text , the same total $g$ factors $g_T^{\rm XX_2}$ and $g_T^{\rm XX_3}$ of the excited biexcitons can be obtained by adding up the individual contributions of conduction and valence band to the overall Zeeman splitting. The expected evolution of the energy levels of the $\sigma^+$-polarized transition of XX$_2$ and XX$_3$ biexcitonic states is schematically depicted in Fig. \ref{S3}(a). The behavior of the $\sigma^-$ transition follows from time-reversal symmetry (not shown here). 

The energy of the bright transition for both excited biexciton states, XX$_2$ and XX$_3$, is not affected by spin (black arrows), as the contributions cancel out. Thus, the $g$ factor $g_{\rm bright}^{\rm X}  = g^{\rm X} \approx -4$ is determined by the orbital contribution from the valence band, see Eq.~(4) of the manuscript. However, the dark states experience a different behavior for XX$_2$ and XX$_3$. In the dark state of XX$_2$, the contributions from spin in the conduction and valence bands evolve anti-parallel. The energy shift of the dark exciton in $\sigma^+$-active XX$_2$ complex with respect to its time-reversal counterpart is described by the $g$ factor
\begin{equation}
g_{\rm dark}^{\rm X} ({\rm XX}_2) = -g_c^{\rm orb} + g_c^{\rm sp} + g_v^{\rm orb} + g_v^{\rm sp} \approx 8\:,
\end{equation}
and therefore $g_T^{\rm XX_2} = g^{\rm X} + g_{\rm dark}^{\rm X} ({\rm XX}_2) \approx 4$.

On the other hand, the contribution from spin in the conduction and valence band evolves parallel in the dark state of XX$_3$, and hence
\begin{equation}
g_{\rm dark}^{\rm X} ({\rm XX}_3) = g_c^{\rm orb} - g_c^{\rm sp} + g_v^{\rm orb} + g_v^{\rm sp} \approx 4\:,
\end{equation}
yielding $g_T^{\rm XX_3} = g^{\rm X} + g_{\rm dark}^{\rm X} ({\rm XX}_3) \approx 0$. The orbital contribution for both dark states is the same. The evolution of the individual $g$ factors of the bright and dark state and the total $g$ factor $g_T^{\rm XX_2}$ of the biexciton of XX$_2$ and XX$_3$  are summarized in Fig. \ref{S3}(b).

\subsection{4.~Role of finite biexciton lifetime}

The analysis performed in the main text demonstrates already good agreement between the experiment and the theory for the biexciton polarization using an estimated total $g$ factor $g_T^{\rm XX_2}=4$ and the effective temperature $T=50$~K. The discrepancy between the temperature inferred from this fit and from Varshni fit $T_L=35$~K in Sec.~{2} of the Supplemental materials can be related with overall inaccuracies in determination of the lattice temperature and the biexciton $g$ factor. Additionally, the finite ratio of the lifetime of the biexciton to its relaxation time between Zeeman-split states, $\tau_0/\tau_s$, results in a depolarization of emission as compared with the thermal one. Qualitatively, this is because biexcitons cannot reach thermal equilibrium during their lifetime. In order to analyze this effect quantitatively we introduce the set of kinetic equations for the occupancies of Zeeman-split states of the XX$_2$ complex, $N_\pm$, emitting in $\sigma^\pm$ circular polarizations respectively:
\begin{subequations}
\label{set}
\begin{align}
\frac{d N_+}{dt} + \frac{N_+}{\tau_0} = \Gamma_{+-} N_- - \Gamma_{-+}N_+ + G/2,\\
\frac{d N_-}{dt} + \frac{N_-}{\tau_0} = \Gamma_{-+} N_+ - \Gamma_{+-}N_- + G/2.
\end{align}
\end{subequations}
Here $G$ is the biexciton generation rate (assumed to be the same for both split states), $\Gamma_{+-}$ ($\Gamma_{-+}$) is the transition rate from $\sigma^-$ to $\sigma^+$ (from $\sigma^+$ to $\sigma^-$) polarized state, whose ratio is given by the Boltzmann factor
\begin{equation}
\frac{\Gamma_{+-}}{\Gamma_{-+}} = \exp{\left(-\frac{g_*^{\rm XX_2}\mu_B B}{k_B T}\right)}.
\end{equation} 
The latter equation assumes that the transitions between the Zeeman-split states are provided by a reservoir at a temperature $T$. By analogy with a two-level system one can introduce the pseudospin $S=N_+-N_-$, which in accordance with Eqs.~\eqref{set} obeys the following kinetic equation
\begin{equation}
\label{S}
\frac{dS}{dt} + \frac{S}{\tau_0} + \frac{S- S_T}{\tau_s} = 0.
\end{equation}
Here, the pseudospin relaxation time $\tau_s = 1/(\Gamma_{+-}+\Gamma_{-+})$, and the equilibrium spin $S_T$ is given by 
\begin{equation}
\label{ST}
S_T = - \frac{1}{2} N \tanh{\left(\frac{g_*^{\rm XX_2}\mu_B B}{2k_B T} \right)},
\end{equation}
with $N=N_++N_-=G\tau_0$ being the total number of biexcitons in the steady state. It follows from Eqs.~\eqref{S} and \eqref{ST} that the steady-state pseudospin of biexcitons differs from the equilibrium one and is given by
\begin{equation}
\label{S0}
S_0 = \frac{\tau_0}{\tau_0+\tau_s} S_T.
\end{equation}
The quantity $f = \tau_0/(\tau_0+\tau_s)$ is known as the dynamic factor accounting for the limitation of the biexciton polarization during its lifetime.

It is noteworthy that in sufficiently small magnetic fields where $|g_*^{\rm XX_2}\mu_B B/ (2k_B T)| \ll 1$ the steady-state pseudospin is given by the equilibrium expression
\begin{equation}
\label{S0:weak}
S_0 \approx - \frac{g_*^{\rm XX_2}\mu_B B}{4k_B T^*} N,
\end{equation}
with effective temperature 
\begin{equation}
\label{T*}
T^*= T \frac{\tau_0+\tau_s}{\tau_0}>T.
\end{equation}
Fixing the temperature $T=35$\,K to conform with the Varshni fit we arrive at the ratio $\tau_s/\tau_0 \approx 0.4$ yielding the biexciton spin relaxation time $\tau_s \approx 30$~ps.

\subsection{5. Estimation of the biexciton decay rate}

Within an excitonic molecule model, the biexciton radiative decay $\Gamma_{0,XX}$ is related to the exciton decay $\Gamma_{0,X}$ as 
\begin{equation}
\Gamma_{0,XX} = \mu \Gamma_{0,X}\:,
\end{equation}
where $\mu$ is a factor, which depends on the wave vector of the center of mass of the biexciton $K$, the interexcitonic separation within the biexciton $a$ and the wave vector of light $q$~\cite{Citrin1994}. A simple estimation $a \approx 4 \div 5$~nm can be extracted from the value of the XX binding energy $E_B^{\rm XX} = 53$~meV and is in line with theoretical calculations of Ref.~\cite{You2015b}. For a wide range of $K$, such as $Ka \lesssim 1$, the factor $\mu \sim (qa)^2 \sim 10^{-2}$ yields a biexciton lifetime $\tau_0\sim 100$~ps and for an exciton a radiative decay time $\tau_{x,0} \sim 1$~ps~\cite{Robert2016}. A precise quantitative study of the exciton and biexciton decay rates is beyond the scope of the present paper.

\end{document}